\documentclass[fleqn,usenatbib]{mnras}

\usepackage{newtxtext,newtxmath}
\usepackage[T1]{fontenc}

\DeclareRobustCommand{\VAN}[3]{#2}
\let\VANthebibliography\thebibliography
\def\thebibliography{\DeclareRobustCommand{\VAN}[3]{##3}\VANthebibliography}

\usepackage{graphicx}	% Including figure files
\usepackage{amsmath}	% Advanced maths commands
\usepackage{color}
\title[post-merger gravitational waves]{On the possibility to detect gravitational waves from post-merger super-massive neutron stars with a kilohertz detector}

\author[Y. K. Chen et al.]{
Yikang Chen$^{1,2}$,
Bin Liu$^{3}$,
Shunke Ai$^{4}$\thanks{E-mail: shunke.ai@whu.edu.cn},
Lin Lan$^{5}$,
He Gao$^{1,2}$\thanks{E-mail: gaohe@bnu.edu.cn},
Yong Yuan$^{4}$,
and Zong-Hong Zhu$^{1,2}$\thanks{E-mail: zhuzh@bnu.edu.cn}
\\
$^{1}$Institute for Frontier in Astronomy and Astrophysics, Beijing Normal University, Beijing 102206, China\\
$^{2}$Department of Astronomy,
Beijing Normal University, Beijing 100875, China\\
$^{3}$Shanghai Astronomical Observatory, Chinese Academy of Sciences, Nandan Road 80, Shanghai 200030, China \\
$^{4}$Department of Astronomy, School of Physics and Technology, Wuhan University, Wuhan 430072, China\\
$^{5}$CAS Key Laboratory of Space Astronomy and Technology, National Astronomical Observatories, Chinese Academy of Sciences, Beĳing 100101, China\\
}

\date{Accepted XXX. Received YYY; in original form ZZZ}

\pubyear{2015}

\begin{document}
\label{firstpage}
\pagerange{\pageref{firstpage}--\pageref{lastpage}}
\maketitle

\begin{abstract}
The detection of a secular post-merger gravitational wave (GW) signal in a binary neutron star (BNS) merger serves as strong evidence for the formation of a long-lived post-merger neutron star (NS), which can help constrain the maximum mass of NSs and differentiate NS equation of states. We specifically focus on the detection of GW emissions from rigidly rotating NSs formed through BNS mergers, using several kilohertz GW detectors that have been designed. We simulate the BNS mergers within the detecting limit of LIGO-Virgo-KARGA O4 and attempt to find out on what fraction the simulated sources may have a detectable secular post-merger GW signal. For kilohertz detectors designed in the same configuration of LIGO A+, we find that the design with peak sensitivity at approximately $2{\rm kHz}$ is most appropriate for such signals. The fraction of sources that have a detectable secular post-merger GW signal would be approximately $0.94\% - 11\%$ when the spindowns of the post-merger rigidly rotating NSs are dominated by GW radiation, while be approximately $0.46\% - 1.6\%$ when the contribution of electromagnetic (EM) radiation to the spin-down processes is non-negligible. We also estimate this fraction based on other well-known proposed kilohertz GW detectors and find that, with advanced design, it can reach approximately $12\% - 45\%$ for the GW-dominated spindown case and $4.7\% - 16\%$ when both the GW and EM radiations are considered.
%The sensitivity of the detectors in the frequency range from $1{\rm kHz}$ to $3{\rm kHz}$ can be significantly improved by increasing the power in arm cavity and applying the technique of opto-mechanical filer, which makes the fraction reaches approximately $12\% - 45\%$ for the GW-dominated spindown case and $4.7\% - 16\%$ when both the GW and EM radiations are considered.
\end{abstract}


%%%%%%%%%%%%%%%%%%%%%%%%%%%%%%%%%%%%%%%%%%%%%%%%%%

%%%%%%%%%%%%%%%%% BODY OF PAPER %%%%%%%%%%%%%%%%%%

\section{Introduction}
Since the initial detection of the gravitational wave (GW) event GW150914, which resulted from a binary-black-hole (BBH) merger \citep[][for summaries]{abbott2016}, a significant number of GW events arising from binary-compact-object mergers have been detected during the first three operational periods of the LIGO-Virgo-KAGRA collaboration \citep{abbott2019,abbott2021,LIGO2021}. However, it is worth noting that only two binary-neutron-star (BNS) mergers (GW170817 \citep{abbott2017} and GW190425 \citep{abbott2020}) have been identified thus far. Additionally, in the case of these BNS merger events, only GWs from the inspiral phase were successfully detected.

The post-merger GW emission from a BNS-merger system depends on the type of central remnant. In theory, there are four possible evolving routes for the central remnant: \citep{rosswog2000,rezzolla2010,rezzolla2011,howell2011,lasky2014,rosswog2014,ravi2014,gao2016,radice2018,ai2020}: 1)it directly collapses into a black hole (BH); 2) it produces a differential-rotating hyper-massive NS (HMNS) that collapses into a BH within seconds; 3) it produces a differentially rotating HMNS that transitions to a rigidly rotating super-massive neutron star (SMNS) after the differential rotation vanishes and eventually collapses into a BH after the SMNS spins down; 4) it produces a stable neutron star (SNS) after the HMNS phase. The timescale of the post-merger GW signal is determined by the lifetime of the post-merger NS. In BNS-merger simulations, if a HMNS is formed, it is expected to emit a significant amount of post-merger GWs dominated by the quadrupolar $f$-mode, with a typical frequency range of approximately 2 kHz to 4 kHz \citep{xing1994, ruffert1996, shibata2000, hotokezaka2013}. In fact, depending on the equation of state (EoS) of neutron stars, the post-merger GW emission can start at around 1 kHz \citep{maione2017}. However, this type of GW emission rapidly damps within $\sim 100{\rm ms}$ due to either the collapse of the HMNS into a BH (route 2) or the formation of a rigidly rotating NS (route 3 and 4) \citep{baiotti2008,kiuchi2009,rezzolla2011,hotokezaka2013,kiuchi2018}. A post-merger rigidly rotating NS can be approximated as a triaxial ellipsoid, and its ellipticity, defined as
\begin{eqnarray}
\epsilon \equiv \frac{|I_1 - I_2|}{I_3},
\end{eqnarray}
where $I_3$ is the moment of inertia about the rotational axis and $I_1$ and $I_2$ are the other two moments of inertia, quantifies its deformation \citep{bonazzola1996, palomba2001, cutler2002}.  This deformation can be induced through various mechanisms, including magnetic-field-induced deformation \citep{bonazzola1996, palomba2001, cutler2002}, bar-mode instability \citep{lai1995, corsi2009}, and r-mode instability \citep{andersson1998, lindblom1998}. Consequently, the long-lived post-merger NS could emit secular GWs with a frequency twice that of the NS's rotation frequency. Efforts have been made to search for the post-merger GW signal associated with GW170817, but no signal related to it has been detected yet. Unfortunately, due to the limited sensitivity of advanced LIGO in the kilohertz (kHz) frequency range, meaningful constraints on the properties of the central remnant cannot be derived \citep{abbott2017b}. 

The nature of the remnant can potentially be inferred from electromagnetic (EM) observations. In the case of a long-lived post-merger remnant with a significant magnetic field, such as a magnetar, several characteristic features may arise \citep[see][for a review]{zhang2018}. One commonly accepted feature indicating the formation of long-lived NS is the so called internal X-ray plateau. This plateau phase is characterized by a nearly-flat light curve that typically lasts for a few hundred seconds and followed by a steep power-law decay with a slope index greater than $3$. This feature is difficult to explain in the framework of an external-shock-driven afterglow but can be naturally powered by a long-lived NS (magnetar) that eventually collapses into a BH during the steep decay phase \citep{rowlinson2010, rowlinson2013, lu2014, lu2015}. For GW170817, although EM counterparts have been observed across a wide range from gamma-ray to radio frequencies, the lack of distinct EM signals and the absence of early X-ray observations make it challenging to determine the formation of a long-lived neutron star. Even if it was believed that a long-lived neutron star was formed in GW170817, its surface magnetic field is expected to be relatively low \citep{ai2018}. Considering the large uncertainties in EM radiation models, confirming the existence of a long-lived neutron star after a binary neutron star merger requires the search for secular post-merger signals using upgraded GW detectors, particularly those with peak sensitivity in the kilohertz range.

In this study, our focus is on the GW signals emitted by long-lived SMNSs or SNSs formed through BNS mergers. We do not delve into specific mechanisms for deformation, but instead treat $\epsilon$ as a general parameter. In section \ref{sec:basic_formula}, we introduce the basic formula for GW emission from a rigidly rotating NS. In Section 3, we present several designs to enhance the sensitivity of GW detectors in kilohertz. In Section 4, we investigate the population properties related to BNS mergers and their central remnants. In Section 5, we discuss the feasibility of detecting this type of GWs using kilohertz GW detectors. Specifically, we aim to find out that, among all the BNS mergers identified by the currently operating GW detector LIGO A+, what fraction would have a detectable secular post-merger GW signal, with different kilohertz detector used. Finally, our conclusions and discussions are presented in Section 5.

\section{Basic formula (Single emitter)}
\label{sec:basic_formula}
Consider a newly formed post-merger rigidly-rotating NS (Hereafter we call it as post-merger NS) with radius $R$, poloidal magnetic field $B_p$, momentum of inertia $I$ and initial spin period $P_0$. Also, suppose the post-merger NS is slightly deformed as an triaxial ellipsoid with ellipticity $\epsilon$. It would spin down via both GW and magnetic dipole radiation, where the spin-down luminosity can be expressed as \citep{shapiro1983,zhang2001}
\begin{eqnarray}
L_{\rm sd} &=& L_{\rm sd,GW} + L_{\rm sd,EM} \nonumber \\
&=& \frac{32 G I^2 \epsilon^2 \Omega^6}{5c^5} + \frac{B_p^2 R^6 \Omega^4}{6c^3}.
\label{eq:Lsd}
\end{eqnarray}
Here $\Omega = 2\pi/P$ represents the angular velocity of the spinning NS, whose evolution can be calculated as $d\Omega / dt = L_{\rm sd} / (I\Omega)$. Using the convention $Q = 10^n Q_n$ for cgs units hereafter, the spin-down timescale for the EM dominated and GW dominated cases can be estimated as
\begin{eqnarray}
   T_{\rm sd,EM} = \frac{E_{\rm rot}}{L_{\rm sd,EM}} = 2.0\times 10^3 I_{\rm 45} R_{\rm s,6}^{-6} B_{\rm p,15}^{-2} P_{\rm i,-3}^2 ~{\rm s}, 
\label{eq: T_EM}
\end{eqnarray}
and 
\begin{eqnarray}
    T_{\rm sd,GW} = \frac{E_{\rm rot}}{L_{\rm sd,GW}}= 9.1 \times 10^3 I_{45}^{-1}P_{i,-3}^{4}\epsilon_{-3}^{-2}~{\rm s},
\label{eq:T_GW}
\end{eqnarray}
respectively, where
$E_{\rm rot} = (1/2)I\Omega_0^2 \approx 2\times 10^{52} I_{45}P_{\rm i,-3}^{-2} {\rm erg}$ is the initial total rotational kinetic energy. 

In our expression, $\epsilon$ determines the strength of the GW radiation, from which the strain of GWs can be derived as
\begin{eqnarray}
h(t) = \frac{4G\Omega^2}{c^4 d} I \epsilon,
\end{eqnarray}
where $d$ is the distance from the source to the observer. In the frequency domain, the characteristic strain could be expressed as 
\begin{eqnarray}
h_c (f) &=& f h(t) \left(\frac{df}{dt}\right)^{-1/2} %\nonumber \\
\label{eq:hc}
\end{eqnarray}
where $f = \Omega / \pi$ = 2/P stands for the frequency of the GWs so that $df/dt = (1/\pi) d\Omega /dt$. From $h_c$ one can calculated the signal-noise-ratio (SNR) for the GW signal with a specific detector as \footnote{We compare the SNR value calculated through this method with that through the standard matched filtering method assuming there is only white noise, and find that they are nearly the same, while the method presented here is more efficient.}
\begin{eqnarray}
{\rm SNR}^2 = \int_{f_{\rm min}}^{f_{\rm max}} \frac{h_c(f)^2}{f^2 S_n(f)} df,
\end{eqnarray}
with $S_n$ representing the sensitivity of the detector. In practical, $f_{\rm max}$ and $f_{\rm min}$ represent the initial and final frequency related to the initial and final spin period of the post-merger NS during the detection. If post-merger NS collapses into a BH during the detection, $f_{\rm min}$ should be derived from the spin period when the collapsing happens. In the GW-dominated case, the EM radiation term in Equation \ref{eq:Lsd} can be ignored, thus the characteristic strain of the GWs in the frequency domain can be further written as 
\begin{eqnarray}
h_c = \left(\frac{5 I G f}{2 c^3 d^2}\right)^{1/2}.
\label{eq:hc_GW}
\end{eqnarray}

Considering a rigidly rotating NS formed right after the differential rotating phase, it is reasonable to assume its initial spin period to be close to the breaking limit (i.e. Keplerian period), which is a function of the NS mass. During the merger, the total baryonic mass of the system is conserved, which can be calculated as 
\begin{eqnarray}
M_{\rm b,rem} = M_{b,1} + M_{b,2} - M_{\rm ej},
\end{eqnarray}
where $M_{b,1}$, $M_{b,2}$ and $M_{\rm b,rem}$ are the baryonic masses for the two pre-merger NSs and the post-merger central remnant. Once the gravitational masses ($M_1$ and $M_2$) for the pre-merger NSs are determined, the baryonic masses can be calculated from a universal relation $M_b = M_g + 0.80M_g^2$\citep[e.g.][]{gao2020}, for the slow-spinning NSs. $M_{\rm ej}$ is the ejected mass during the merger, which can be obtained from the EM follow-up observation or the numerical simulations for BNS mergers. We adopt $M_{\rm ej} = 0.01M_{\odot}$ in the whole paper \footnote{For GW170817, the total ejected mass is $\sim 0.06 M_{\odot}$ \citep[][and the references therein]{metzger2017}. However, this is fitted in the scenario of a typical r-process kilonova. If a long-lived NS was produced, extra energy would be injected into the ejecta \citep{yu2013,ai2022}, so that the fitted ejecta mass would be modified. Besides, the ejecta mass almost does not change our final results. Therefore, we assume $M_{\rm ej} = 0.01M_{\odot}$ just to make the ejecta mass in a reasonable order of magnitude.}. Considering that the post-merger NS are rapidly rotating, the relation of $M$ and $M_b$ for the slow-spinning NSs is no longer applicable. For a NS spins at the Keplerian period, a new relation $M_b = M + 0.064M_g^2$ would apply, from which the initial gravitational mass for the central remnant ($M_{\rm rem,0}$) can be estimated. Further, one can estimate the initial spin period from another universal relation \citep{ai2020}
\begin{eqnarray}
{\cal P}_0 = -2.697\left(\frac{M}{M_{\rm TOV}}\right)^2 + 4.355 \left(\frac{M}{M_{\rm TOV}}\right) - 0.303,
\end{eqnarray}
where ${\cal P}_0 = P_0 / P_{\rm k,min}$ is the normalized initial spin period. Considering a NS with arbitrary spin period, a more general relation for the baryonic and gravitational masses, which reads as
\begin{eqnarray}
M_b = M + R_{1.4}^{-1} e^{-\frac{1}{4{\cal P}}} \times M^2,
\label{eq:Mb_M_general}
\end{eqnarray}
can be used to convert $M_{\rm b,rem}$ back to the gravitational mass of the central remnant ($M_{\rm rem}$). $R_{1.4}$ is the radius in ${\rm km}$ for the non-rotating NS when $M_{\rm NS} = 1.4M_{\odot}$. ${\cal P} = P/P_{\rm k,min}$ is defined as the normalized NS spin period, where $P_{\rm k,min}$ represents the minimum period of a rigid rotating NS before breaking up. In this work, $R_{1.4} = 12.38{\rm km}$ and $P_{\rm k,min} = 5.85 \times 10^{-4}{\rm s}$ are adopted\footnote{These are the mean values of $R_{1.4}$ and $P_{\rm k,min}$ for the NS EoSs listed in \cite{ai2020}}. Following the spin-down history of the post-merger NS, together with Equation \ref{eq:Mb_M_general}, one can calculate the its spin period as well as the gravitational mass at any time. 
When the criterion \citep{ai2020}
\begin{eqnarray}
{\rm log}_{10}\left(\frac{M_{\rm rem}}{M_{\rm TOV}}-1\right) > -2.740 \times {\rm log}_{10} {\cal P} + {\rm log}_{10} (0.201)
\end{eqnarray}
is satisfied, the massive NS should collapse into a BH. Therefore, the collapsing time ($T_{\rm col}$) and the collapsing spin period ($P_{\rm col}$) can be read out. If the collapsing time is smaller than the length of effective GW data ($T_{\rm d}$), we use $P_{\rm col}$ to calculate $f_{\rm min}$. Otherwise, we should follow the spin-down history until $T_{\rm d}$, and use the spin period at that time to calculate $f_{\rm min}$.

Examples for the evolution of $h_c$ are shown in Figure \ref{fig:Sensitivity}, with $T_{\rm d} \sim 5 \times 10^3 {\rm s}$ assumed as the length of effective GW data for our aimed signal. When magnetic dipole radiation has contribution to the spin-down process, even though with the same $\epsilon$ value, the characteristic strain ($h_c$) of the GW signal would be smaller than the GW-dominated case. If a longer $T_d$ is assumed, the trajectory of $h_c$ would extend to lower frequency, thus a higher SNR is expected.

\section{kilohertz GW detectors} 
\label{sec:detector}
Previous calculations indicate that, the GWs from the post-merger NSs are mainly in kilohertz.  In this section, targeting on this type of GW sources, we propose several kHz GW detectors, with central frequency at 1\,kHz, 2\,kHz, and 3\,kHz, respectively, for comparison. Usually, extending the length of the signal recycling cavity is the most convenient and efficient way to shift the detection frequency band from low to high, which is the main approach used in our designs.

\begin{figure}
	\includegraphics[width=\columnwidth]{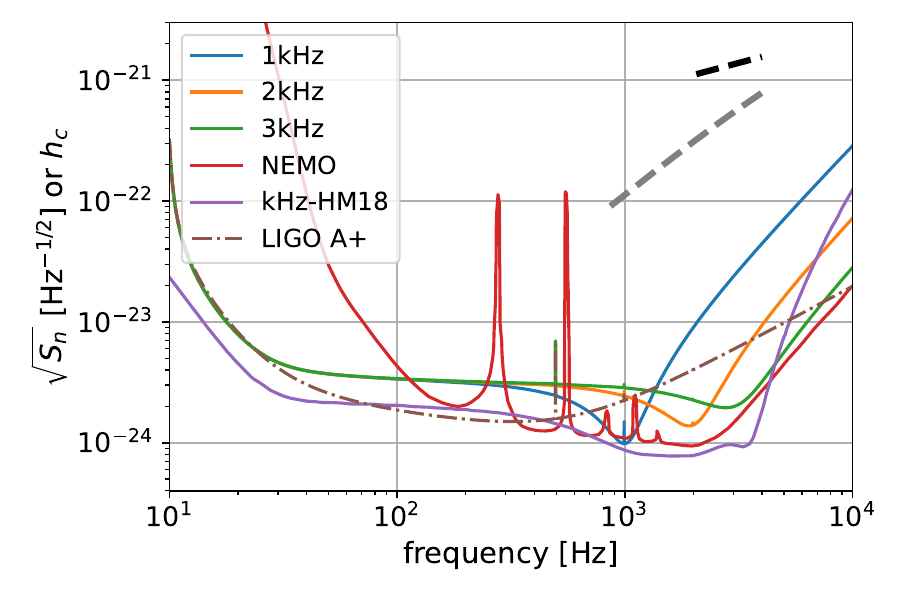}
    \caption{The sensitivity curves for the designed kilohertz GW detectors (solid lines) and the trajectories for $h_c$ evolution (dashed lines) in the frequency domain. As comparison, the sensitivity curve for LIGO A+ is also shown. The red and purple solid lines are extracted from \protect\cite{nemo} and \protect\cite{miao2018}, respectively. Other solid lines are generated by using the python module \emph{pygwinc} \citep{rollins2020}. We adopt the parameters for the post-merger NS as $P_0 = 0.5{\rm ms}$, $I = 1.5\times 10^{45}{\rm g~cm^2}$, $R = 1.2\times 10^5{\rm cm}$. Assume the post-merger NS will never collapse and locate at $40$Mpc from the observer. For the black dashed line, $\epsilon = 10^{-2}$ and $B_p = 0$ are assumed. For the grey dashed line,  $\epsilon = 10^{-2}$ and $B_p = 10^{15}{\rm G}$ are adopted.}   
    \label{fig:Sensitivity}
\end{figure}

In this work, we design kHz detectors based on LIGO A+. Considering that the target detecting frequency band is far away from the free spectrum range of the configuration of the kHz detector, the influence of antenna response can be ignored.
The arm length are short enough that the gravitational wave strain can be approximately treated as static \citep{Reed2018}. The signal recycling cavity (SRC) and arm cavity can be considered as coupled cavities. By decreasing the transmission of the signal recycling mirror, the wide band can be changed to narrow band. Meanwhile, the sensitivity around the resonant frequency will become better. In this case, the sum of the phase shift reflected from the arm cavity $\phi_{\rm arm}$ and the phase shift of light through SRC $\phi_{\rm src}$ need to achieve resonance conditions, which can be described as \citep{miao2018,Denis2019} 

\begin{eqnarray}
\phi_{\rm arm}(\Omega)+\phi_{\rm src}(\Omega) = \pi.
\label{eq2}
\end{eqnarray}
$\phi_{\rm arm}$ and $\phi_{\rm src}$ are calculated as
\begin{eqnarray}
\phi_{\rm src}(\Omega)=\frac{2\Omega L_{\rm src}}{c}
\end{eqnarray}
and
\begin{eqnarray}
\phi_{\rm arm}(\Omega)={\rm atan} \frac{\Omega }{\gamma_{\rm arm}}
\end{eqnarray}
with
\begin{eqnarray}
\gamma_{\rm arm}=\frac{c T_{\rm arm}}{4 L_{\rm arm}},
\end{eqnarray}
where $L_{\rm src}$ is the length of SRC. $\Omega=2\pi f$ is the sideband frequency. $\gamma_{\rm arm}$ is the arm cavity bandwidth. $L_{\rm arm}$ is the length of arm cavity. $T_{\rm arm}$ is the transmission of the arm cavity.  We adopt the same arm cavity configuration of LIGO A+ \citep{Ligo+}. $L_{\rm src}$ and sideband frequency $f$ have a corresponding relationship. The transmission of SRM $T_{\rm src}$ determines the width of the sideband of the detector's power spectral density. The parameters of the detectors designed in this paper is shown in Table 1.
Meanwhile, the sensitivity curves for our designed detectors, with power spectrum density as the vertical axis, are shown in Figure \ref{fig:Sensitivity}.

Moreover, by increasing the power within the arm cavity and implementing an opto-mechanical filter, one can significantly amplify the signal, resulting in an approximately five-fold increase in the SNR \citep{miao2018}. To achieve the same goal by simply increasing the light intensity, it needs to be increased by about 25 times, which is extremely hard due to the difficulties in controlling the input laser power, dealing with the mirror absorption thermal effect and the dynamic instability. Although the opto-mechanical filter on GW detectors is also a raw technique right now, designs have been made in the literature \citep[e.g.][]{miao2018}, which can increase the sensitivity of the detector in a much wider frequency range, totally covering the peak frequencies for our designed kHz detector. We further consider the design of the Neutron Star Extreme Matter Observatory (NEMO) \citep{nemo}, which is a proposed $2.5$-generation GW detector to be constructed in Australia with peak frequency in kilohertz band. NEMO has higher arm circulating power ($4.5$ MW), lower longer laser wavelength ($2~{\rm \mu m}$) and advanced cooling technique compared to our designs. Their sensitivity curves are also shown in Figure \ref{fig:Sensitivity}.

\begin{table}
\caption{Parameters of different kHz detector designed based on LIGO A+}\label{ta:par}
\begin{tabular}{cccc}
    Parameters &  & value & \\
    \hline\\
    Arm length, $ L_{\rm arm}$ &  & 4\,km & \\
    Arm power, $ P_{\rm arm}$  &  & 750\,kW & \\
    Arm transmission, $ T_{\rm arm}$ & & 0.014 & \\'   
    SRC transmission, $ T_{\rm src}$ & & 0.03 &  \\
    Injected squeezing & &12\,dB & \\
    Injection loss & &5\% & \\
    Readout loss & &10\% & \\
    Filter cavity length &  & 300\,m &  \\
    Filter cavity loss &  & 60\,ppm & \\
    center frequency & 1\,kHz  & 2\,kHz  & 3\,kHz \\
    SRC length, $ L_{\rm src}$  & 2000\,m & 500\,m & 200\,m \\
               &  &  & \\
    \hline
                    
\end{tabular}
\label{tab:detector}
\end{table}

\section{Population Properties of the sources}
\subsection{distribution of redshift}
\label{sec:redshift_dis}
The red-shift distribution of BNS mergers is determined by their merger rate at each red-shift. Usually, the BNS merger rate in unit volume follows the star formation rate (SFR) convoluting with the a time delay from the formation of the BNS system to the merger \citep{wanderman2015,sun2015}, which reads as
\begin{eqnarray}
R_{\rm merger} (z) = R_{\rm merger,0} \int_0^z f(z^{\prime}) G(\tau) \frac{d\tau}{dz} dz^{\prime},
\end{eqnarray}
where the $f(z)$ = {\rm SFR}(z) / {\rm SFR}(z=0) is the normalized SFR, with the SFR fitted as \citep{yuksel2008}
\begin{eqnarray}
   {\rm SFR}(z) \propto \left[(1+z)^{3.4 \eta} + \left(\frac{1+z}{3000}\right)^{-0.3\eta} + \left(\frac{1+z}{9}\right)^{-3.5\eta} \right]^{1/\eta},
\end{eqnarray}
where $\eta = -10$ is adopted. $\tau$ stands for the time delay and $G(\tau)$ describes its probability density function. Assume the time delay follows a log-normal distribution \citep{wanderman2015,sun2015}, which reads as
\begin{eqnarray}
G(\tau) = \frac{1}{\sqrt{2\pi} \sigma_{\tau} \tau} {\rm exp} \left(\frac{{\rm ln}\tau - {\rm ln}{\tau_d}}{2\sigma_{\rm tau}^2}\right),
\end{eqnarray}
where $\tau_d = 2.9~{\rm Gyr}$ and $\sigma_\tau = 0.2~{\rm ln}({\rm Gyr})$. 

The BNS merger rate in a certain rad-shift bin can finally be calcualted as
\begin{eqnarray}
d\dot{N}(z) = \frac{R_{\rm merger}(z)}{1+z} \frac{dV}{dz} dz,
\label{eq:Nz}
\end{eqnarray}
where $V$ is the comoving volume of the universe and
\begin{eqnarray}
\frac{d V(z)}{d z}=& 4 \pi\left(\frac{c}{H_0}\right)^3\left(\int_0^z \frac{d z}{\sqrt{1-\Omega_m+\Omega_m(1+z)^3}}\right)^2 \nonumber\\ & \times \frac{1}{\sqrt{1-\Omega_m+\Omega_m(1+z)^3}}.
\end{eqnarray}
with $H_0 = 68.7{\rm km~s^{-1}~Mpc^{-1}}$ and $\Omega_m = 0.308$ adopted \citep{Planck_Collaboration2016}. 

\subsection{distribution of mass}
\label{sec:mass_dis}
The mass of the merger remnant can be calculated from the masses of the two pre-merger NSs. Before the detection of GW190425, the mass distribution for BNS mergers are supposed to follow that for the Galactic double NS, which can be described by a Gaussian distribution with central value $\mu = 1.33$ and standard deviation $\sigma = 0.11$ \citep{kiziltan2013}. However, the total mass of the BNS-merger system GW190425 reaches $M_{\rm tot} = 3.4M_{\odot}$ \citep{abbott2020}, which exceeds the $5\sigma$ limit of the Gaussian distribution presented above, indicating the BNS merger systems may divert from that for the Galactic double NSs.

In this work, since the correct distribution of BNSs are still under debate, we still use the mass distribution for Galactic double NSs to simulate all the BNS merger systems. This treatment will make our estimate on the possibility to detect the secular post-merger GW signals optimal, due to more rigidly rotating NSs would be formed. To simulate the merger remnants, another crucial physical quantity is $M_{\rm TOV}$. Recently,  a millisecond pulsar with mass $M = 2.35 \pm 0.17$ has been found \citep{romani2022}, indicating that very likely $M_{\rm TOV} \gtrsim 2.35 M_{\odot}$. Combined with the previous constraints on $M_{\rm TOV}$ based on the generally believed scenario that the merger remnant of GW170817 collapsed into a BH, e.g. $M_{\rm TOV} \lesssim 2.3M_{\odot}$ \citep{shibata2019}, we adopt $M_{\rm TOV} = 2.35M_{\odot}$. However, the constraints on $M_{\rm TOV}$ from GW170817 is not necessarily correct because the type of merger remnant are not determined, thus the value of $M_{\rm TOV}$ could be higher. The value adopt here may offset some overestimation of the production of rigidly rotating NSs as the merger remnants introduced by the mass distribution assumed above.

\subsection{distributions of $\epsilon$ and $B_p$}
\label{sec:epsilon_B_dis}
The distribution of $\epsilon$ and $B_p$ can be inferred from the breaking times of the X-ray internal plateaus. The SGRBs followed by a X-ray internal plateau as well as the breaking times are listed in Table A1 (See Appendix \ref{sec:breaking times}). 

%From the mass distribution for the BNS-merger systems, along with the method to calculate calculate the remnant mass, the initial condition for each SMNS is known, including the initial gravitational mass and period. Then, assuming a certain set of $\epsilon$ and $B_p$, a theoretical distribution of the collapsing time can be obtained. Then, make a Kolmogorov-Smirnov (KS)  test between the observed X-ray plateau breaking times and the calculated SMNS collapsing times to test if they come from the same distribution. 

\begin{figure}
	\includegraphics[width=\columnwidth]{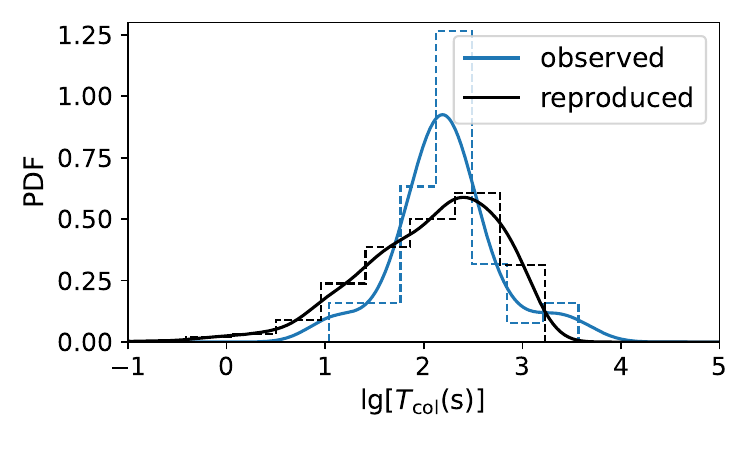} \\
    \includegraphics[width=\columnwidth]{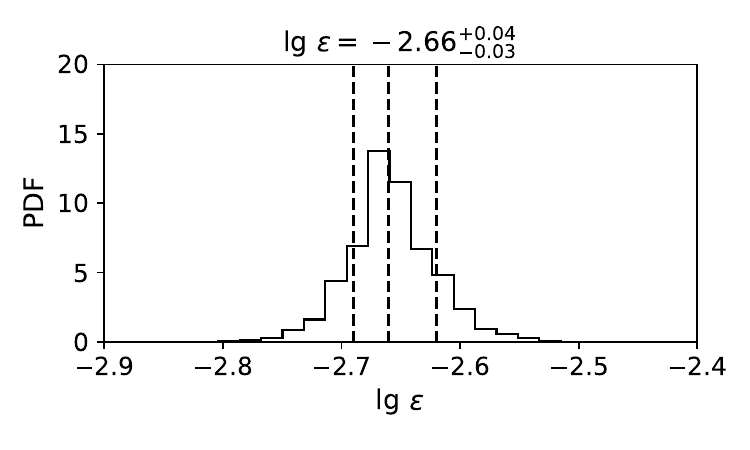}
    \caption{The distributions of collapsing time of SMNSs are shown in the upper panel while the distribution of $\epsilon$ is shown in the lower panel. In the upper panel, the solid lines represent the fitted probability density function with kernal density estimation {\bf using the python module \emph{scipy.stats.gaussian\underline{~}kde}}. The black line and histogram represent the collapsing times reproduced from the $\epsilon$ distribution in the lower panel, considering the mass distribution of BNS mergers and $M_{\rm TOV}$.}   
    \label{fig:Tcol_epsilon}
\end{figure}

First of all, we assume the magnetic field on all of the post-merger NSs are efficiently small so that the spin-down processes are dominated by GW radiation. In this scenario, $\epsilon$ becomes the only free parameter. According to the Bayesian theory, the posterior distribution of $\epsilon$ can be expressed as
\begin{eqnarray}
p(\epsilon~|~T_{\rm b,obs}) \propto {\cal L}(T_{\rm b,obs}~|~\epsilon) p(\epsilon) 
\label{eq:eps_posterior}
\end{eqnarray}
where $p(\epsilon)$ represent the prior distribution of $\epsilon$ and ${\cal L}(T_{\rm b,obs}~|~\epsilon)$ represents the likelihood of a certain $\epsilon$. Instead of an analytical formula, the likelihood is obtained by conducting Monte Carlo simulation. For each value of $\epsilon$, we simulate $10^3$ BNS-merger sources according to the methods introduced in Section \ref{sec:basic_formula} and Section \ref{sec:epsilon_B_dis}. For the sources where a long-lived super-massive neutron star is produced, the corresponding collapsing time can also be calculated. We compare the simulated collapsing time samples with the observed breaking times of X-ray internal plateaus, by conducting a Kolmogorov-Smirnov (KS) test using the python module \emph{scipy.stats.kstest}, to check if they follow the same distribution. The p value of the KS test represents the probability for the two datasets coming from the same distribution. And we assume the likelihood in Equation \ref{eq:eps_posterior} is proportional to this p value. Here, we adopt a uniform prior distribution for $\epsilon$ from $10^{-5}$ to $10^{-1}$ in logarithmic space. Then, we run the Markov Chain Monte Carlo (MCMC) simulation based on Equation \ref{eq:eps_posterior}, using the python module \emph{emcee} \citep{Foreman-Mackey2013}, to obtain a sample of $\epsilon$.
%The likelihood is proportional to the p value of the KS test, which has the physical meaning that the probability for the two datasets coming from the same distribution. 
The distribution of collapsing times of SMNSs and the resultant distribution of $\epsilon$ from MCMC simulation are shown in Figure \ref{fig:Tcol_epsilon}. Here ${\rm lg}(\epsilon) = -2.66^{+0.04}_{-0.03}$ with the uncertainty at $1\sigma$ confident level is obtained. Using the inferred $\epsilon$ distribution, based on the mass distribution of BNS mergers and $M_{\rm TOV}$, we generate a set of collapsing times, whose distribution is shown with the black line and histogram in the upper panel of Figure \ref{fig:Tcol_epsilon}. The reproduced collapsing time distribution is generally consistent with the distribution of the observed ones, which is shown with the blue line and histogram.

\begin{figure}
	\includegraphics[width=\columnwidth]{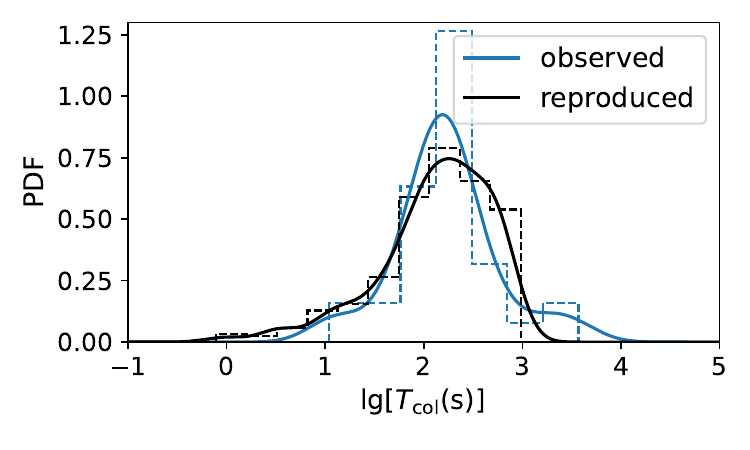} \\
    \includegraphics[width=\columnwidth]{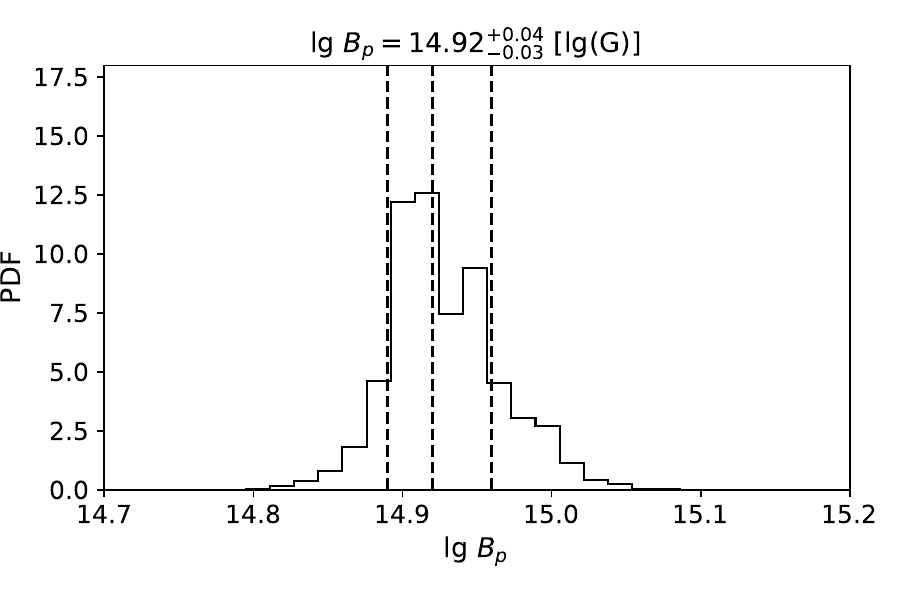}
    \caption{Similar as figure \ref{fig:Tcol_epsilon}, but EM-dominated spindown is assumed.} 
    \label{fig:Tcol_B}
\end{figure}

Next, as the intermediate step, we assume the spin-down processes for all the post-merger rigidly rotating NSs are dominated by magnetic dipole radiation. In this scenario, $B_p$ becomes the only free parameter. We adopt a uniform prior distribution for $B_p$ from $10^{13}{\rm G}$ to $10^{17}{\rm G}$ in logarithmic space and follow the procedure above to find the posterior distribution of $B_p$. The result is is shown in the upper panel of Figure \ref{fig:Tcol_B}, while The resultant collapsing time distribution for simulated SMNSs, and the observed distribution of the breaking times of X-ray internal plateaus, are shown in the lower panel.

\begin{figure}
     \includegraphics[width=\columnwidth]{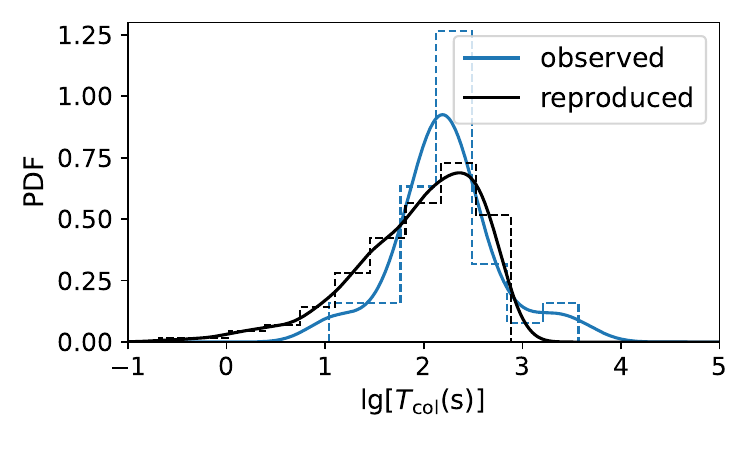} \\
	\includegraphics[width=\columnwidth]{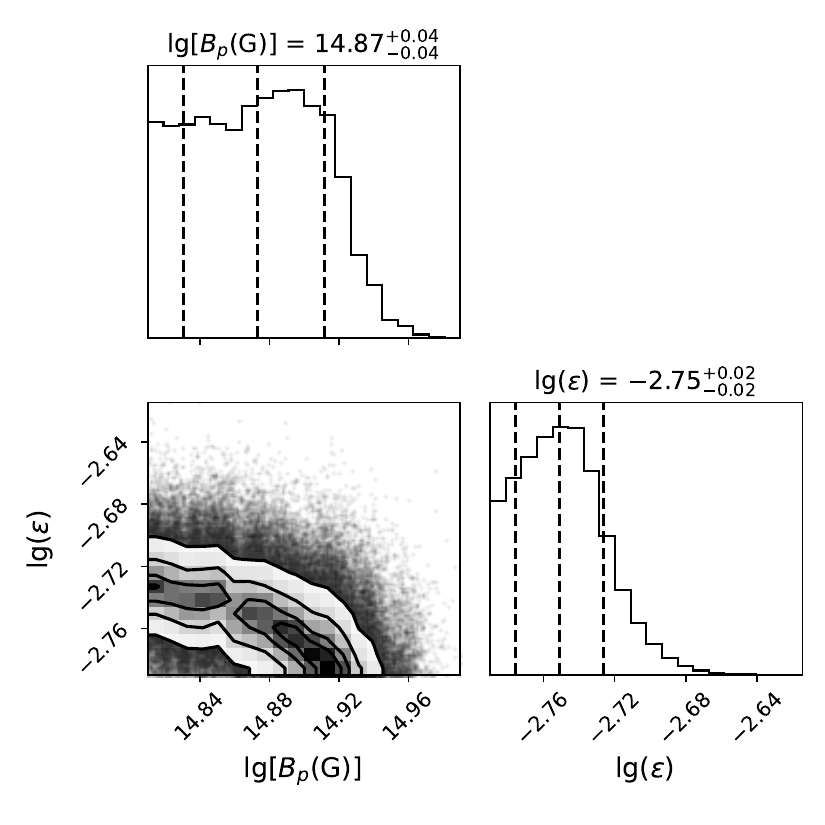}
    \caption{Similar as Figure \ref{fig:Tcol_epsilon} and \ref{fig:Tcol_B}, the distribution of collapsing time of SMNSs are shown in the upper panel while the joint distribution of $\epsilon$ and $B_p$ is shown in the lower panel.} 
    \label{fig:Tcol_epsilon_B}
\end{figure}

For a rigidly rotating NS, if its triaxial shape is induced by the magnetic field, some theories suggest a relation $\epsilon \propto B_p^2$ \citep{bonazzola1996,haskell2008}. However, since the origin of $\epsilon$ is unclear, to be conservative, we treat $\epsilon$ and $B_p$ as independent. When both GW and EM radiation have contributions to the spin-down process, the range of $\epsilon$ and $B_p$ should in the range constrained in Figure \ref{fig:Tcol_epsilon} and \ref{fig:Tcol_B}, otherwise, due to they are independent, collapsing times outside the range of the observed ones would be generated. Here, we use the $3\sigma$ limits ($99.73\%$) of the two parameters as their boundaries (${\rm lg}(\epsilon) \in [-2.79,-2.52]$ and ${\rm lg}(B_p) \in [14.81,15.07]$). The joint posterior distribution of $\epsilon$ and $B_p$ can be written as
\begin{eqnarray}
p(\epsilon,B_p~|~T_{\rm b,obs}) \propto {\cal L}(T_{\rm b,obs}~|~\epsilon,B_p) p(\epsilon) p(B_p),
\label{eq:eps_B_posterior}
\end{eqnarray}
while again the likelihood is calucated by Monte Carlo simulation. The only difference compared to the procedure describe above is that we simulate the BNS mergers to find their corresponding the collapsing times under each set of $\epsilon$ and $B_p$. The joint constraint on $\epsilon$ and $B_p$ is shown in the upper panel of Figure \ref{fig:Tcol_epsilon_B}. From the lower panel, it can be seen that the generated collapsing times have a distribution being consistent with that for the observed ones.

\section{Detectability}
Knowing basic formula for a single emitter and the population properties for the sources, we can estimate their the detectability with a specific GW detector. Here we focus on the BNS mergers that can be detected during LIGO-Virgo-KAGRA O4, which has a detecting limit as $170{\rm Mpc}$ \citep{abbott2020b}, and try to find out on what fraction the identified BNS mergers may followed by a detectable secular post-merger GW emission. 

We randomly generate $5\times 10^5$ BNS merger systems within $170{\rm Mpc}$, according to the red-shift distribution introduced in Section \ref{sec:redshift_dis}. In principle, our simulated sources are all at the near universe where the event rate density, $R_{\rm merger}$, can be treated as constant over red-shift. Although the method to calculate the red-shift distribution introduced above is overqualified, we still use it to be general. For each system, the masses of pre-merger NSs are randomly set, according to the mass distribution presented in \ref{sec:mass_dis}. Then we could judge whether a rigidly rotating NS is formed after the merger, based on the criterion $M_{\rm rem,0} < 1.2M_{\rm TOV}$ \citep{cook1994,lasota1996,breu2016,ai2020}. Once a post-merger rigidly rotating NS is produced, we randomly generate $\epsilon$ and $B_p$ according to their inferred distributions presented in Section \ref{sec:epsilon_B_dis}. Next, we follow its spin-down history and calculate the characteristic strain for the GW radiation and its SNR with a specific GW detector. Setting a threshold SNR, for each detector, we can count the number of sources with a detectable post-merger GW signal and record the fraction.

\subsection{GW-dominated spindown}
\label{sec:GW_dominated}

\begin{figure*}
    \centering
    \begin{tabular}{ll}
    \resizebox{80mm}{!}{\includegraphics{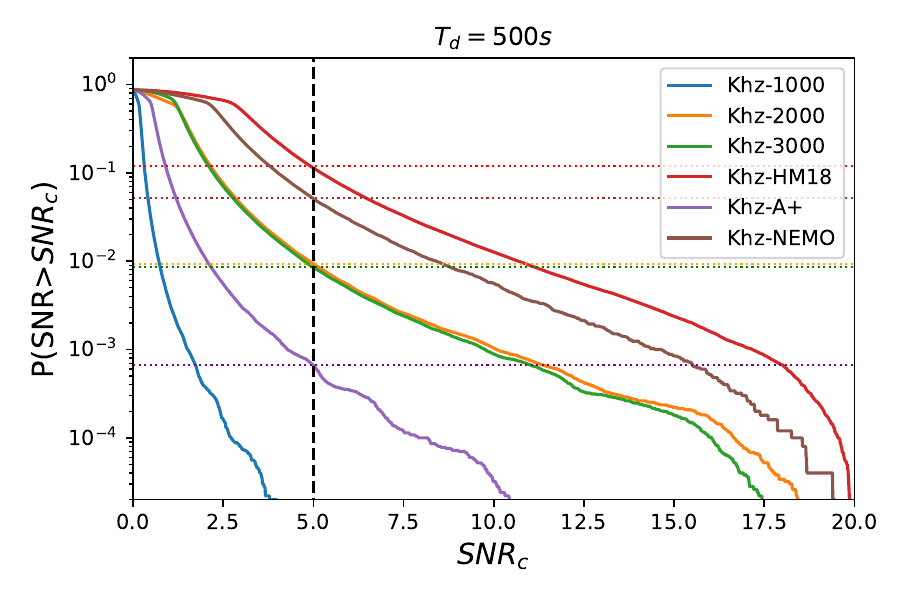}} &
    \resizebox{80mm}{!}{\includegraphics{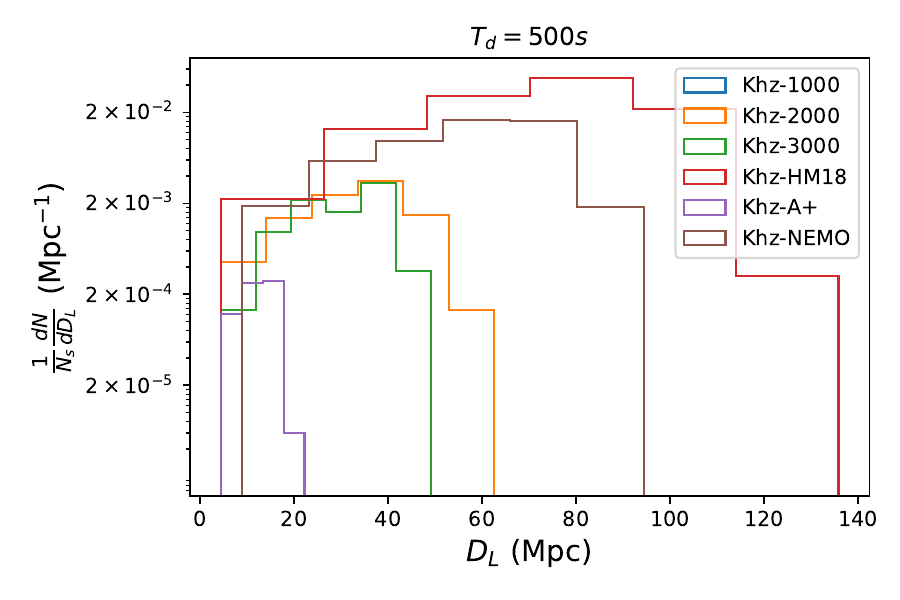}}
    \\
    \resizebox{80mm}{!}{\includegraphics{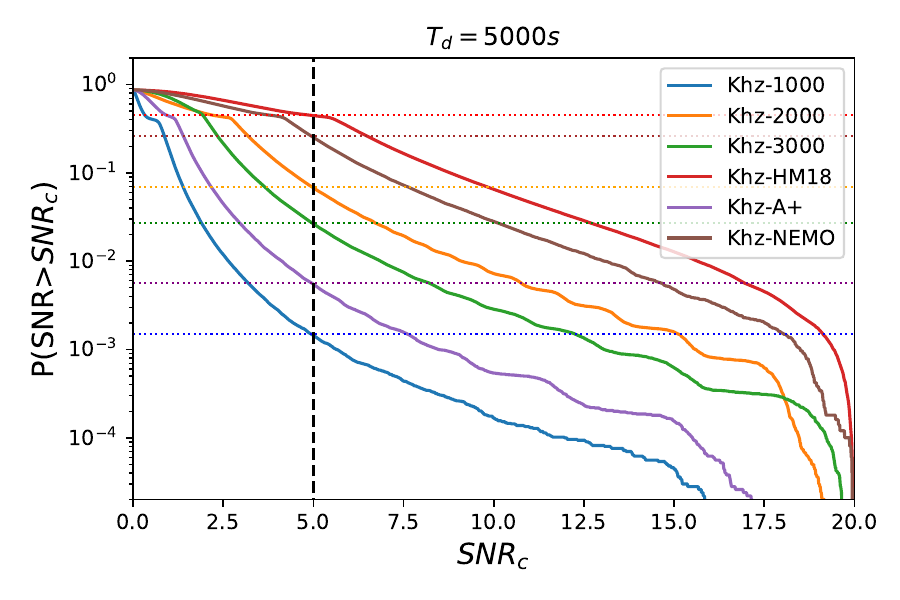}} &
    \resizebox{80mm}{!}{\includegraphics{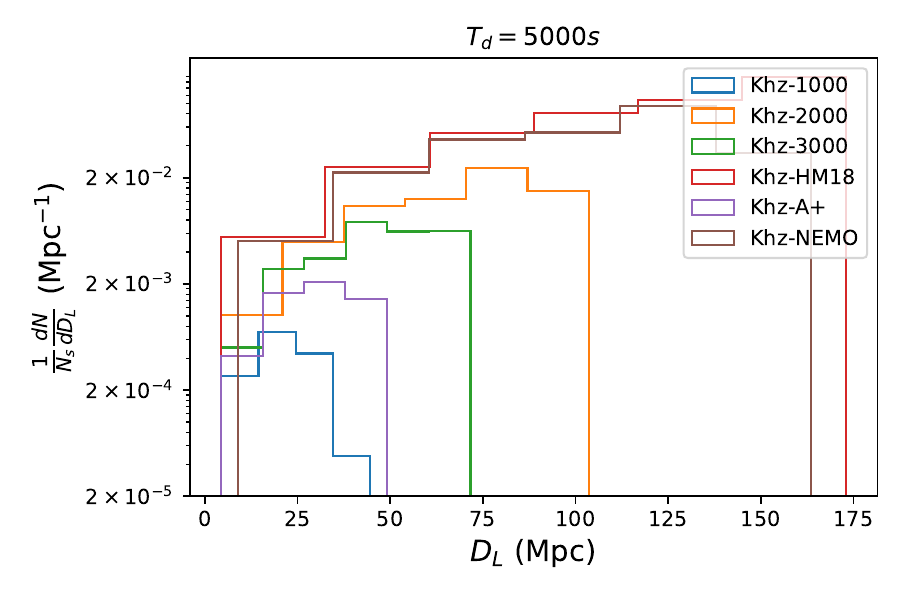}}
    \\
    \resizebox{80mm}{!}{\includegraphics{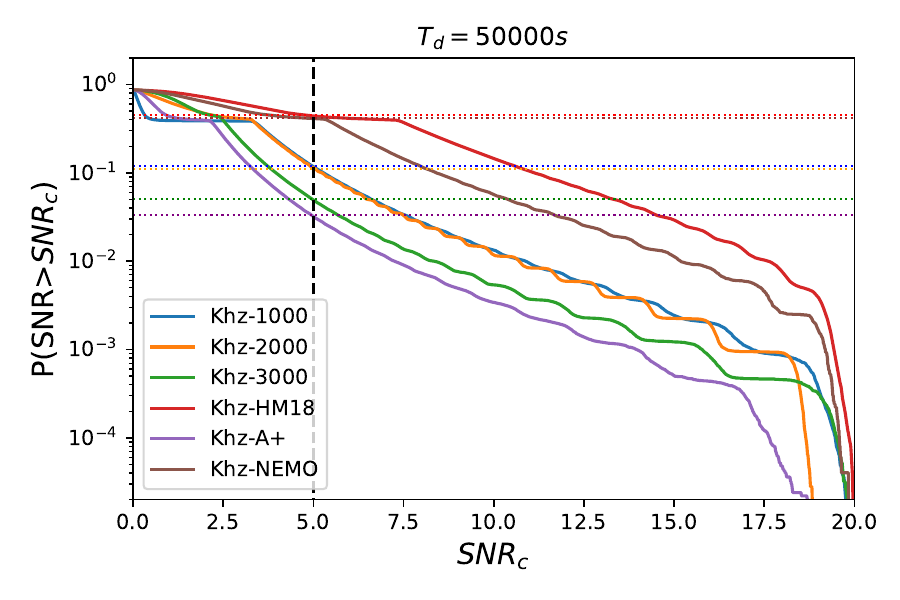}} &
    \resizebox{80mm}{!}{\includegraphics{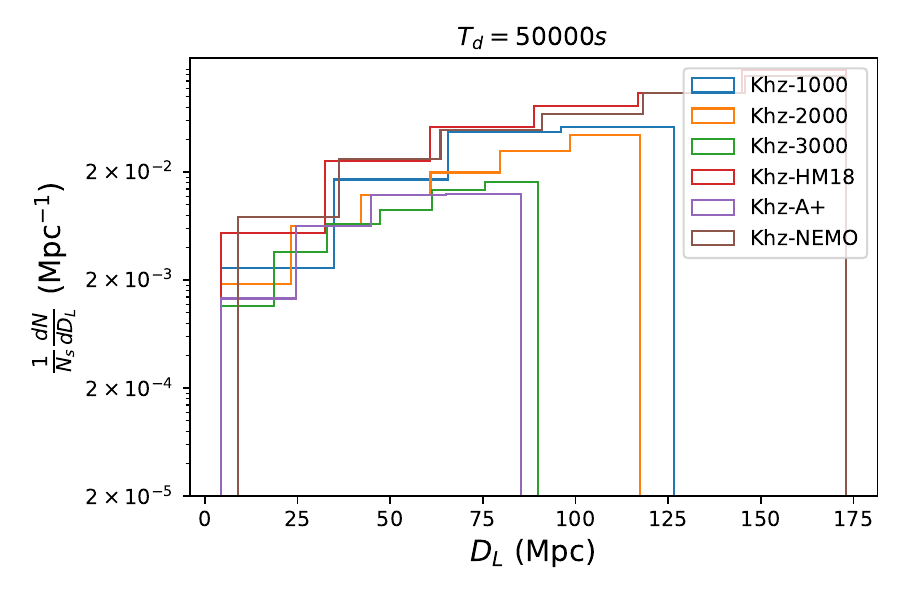}}
    \end{tabular}
    \caption{The left panels show the cumulative SNR distribution for the simulated post-merger GW signals, while the right panels show the distribution of the distances from the GW sources to the observer for those with ${\rm SNR} > 5$. The upper, middle and lower panels show the cases that $T_d = 5 \times 10^2s$, $5 \times 10^3s$, and $5 \times 10^4s$, respectively. Different colors stand for different GW detectors applied. All of these panels is in the case that the spin-down process is dominated by the GW radiation.} 
    \label{fig:SNR and D_GW_dominate}
\end{figure*}

For different kilohertz GW detectors, the cumulative distribution for SNRs of secular post-merger GWs are shown in the left panels of Figure \ref{fig:SNR and D_GW_dominate}. We also show the distribution of SNR when LIGO A+ is used to search for the secular post-merger GWs. Assuming ${\rm SNR} > 5$ as the criterion, the fraction of the simulated sources that have a detectable secular post-merger GW signal can be read. The values are listed in Table 2. 
 
When the length of valid data ($T_d$) is relatively short, the GW detectors with peak sensitivity at a higher frequency has better performance. This is because, with a short $T_d$, at the end of detection, most of the newly formed rigidly rotating NSs have not spun down yet, in which case the evolving trajectory of $h_c$ would only span a small range of frequency so that the integral SNR remains relatively small. As the length of valid data ($T_d$) increases, for all the detectors, the fraction of sources with detectable post-merger GW signals would increase. This is also reasonable, because the evolving trajectory of $h_c$ for some long-lived NSs would extend to lower frequencies along with the spin-down processes, resulting in the increases of the integral SNRs, especially when GW detectors peaks at lower frequencies are used. This explains why the detector with peak sensitivity at $3000{\rm Hz}$ is much better than that peaks at $1000{\rm Hz}$ when $T_d = 5 \times 10^2 {\rm s}$, while it becomes opposite when $T_d = 5 \times 10^4{\rm s}$. Although for most of the cases, increasing the data length $T_d$ can effectively lead to a higher fraction of sources that have detectable post-merger GW signals, a saturated fraction might exist. For example, using the detector proposed by \cite{miao2018}, this fraction no longer increases when $T_d$ increases from $5 \times 10^3{\rm s}$ to $\times 10^4{\rm s}$. This is because, for any source within $170{\rm Mpc}$, if it can survive for $5 \times 10^3{\rm s}$ and has a $\epsilon$ value within the inferred range, it must have reached threshold SNR at an earlier time. The histogram of the distances of all the sources with a detectable secular post-merger GW signal are shown in the right panels of Figure \ref{fig:SNR and D_GW_dominate}. The vertical axes are normalized with the total number of simulated sources. As can be seen, for the secular post-merger GW signals, the detecting rate might not necessarily be proportional to the cube of the detecting limit of a detector, because the detecting limit is determine by the GWs from SNSs, while some high-frequency detectors might also sensitive to the sources collapsing at earlier times.

The performances of the kilohertz GW detectors designed basically with the same technique used in LIGO A+ are not satisfactory. Even when $T_d = 5\times 10^4{\rm s}$, only $3\% - 10\%$ BNS sources identified in LIGO-Virgo-KAGRA O4 can have a detectable post-merger GW signal. When $T_d = 5 \times 10^2{\rm s}$ is assumed, this fraction decreases to less than $1\%$. However, they have been much better than when LIGO A+ is used to search for the secular post-merger GW signals. Among them, the performance of the detector peaks at $2000{\rm Hz}$ is the best and the most stable. 

Using NEMO, the fraction of BNS-merger events with a secular post-merger signal would be much higher. For example, the fraction increases from $\sim 5\%$ when $T_d = 5 \times 10^2{\rm s}$ to $\sim 42\%$ when $T_d = 5 \times 10^4{\rm s}$. If more quantum technologies were applied, i.e. using the design in \cite{miao2018}, this fraction could reach $\sim 11\%$ when $T_d = 5 \times 10^2{\rm s}$ and approaches a saturated value as $\sim 45\%$ when $T_d > 5 \times 10^3{\rm s}$.

\begin{table}                                          
\begin{center}{\scriptsize                                          \caption{The fraction of BNS mergers, inside the detecting limit of LIGO-Virgo-KAGRA O4 ($170~{\rm Mpc}$), with a detectable post-merger GW signal. The spin-down process is assumed to be dominated by GW radiation.}

\begin{tabular}{llll}   
\hline
\hline
&$T_d = 5 \times 10^2~{\rm s}$ &$T_d = 5 \times 10^3~{\rm s}$ & $T_d = 5 \times 10^4~{\rm s}$  \\
\hline
LIGO A+ &0.067\%& 0.56\%&3.3\%  \\
kHz-1000 & 0.00060\%&0.15\% &12\%  \\
kHz-2000 &0.94\%&6.9\%&11\%    \\
kHz-3000 &0.86\%&2.7\%&5.0\%   \\
NEMO &5.2\%&26\%&42\%    \\
kHz-HM18 &12\%&45\%&45\%    \\
\hline
\hline
\end{tabular} 
\\ \hspace*{\fill} \\

}
\end{center}                                         \label{table:GW_dominated} 
\end{table}

\subsection{The effect of dipole magnetic field}
\label{sec:GW_EM}

\begin{figure*}
    \centering
    \begin{tabular}{ll}
    \resizebox{80mm}{!}{\includegraphics{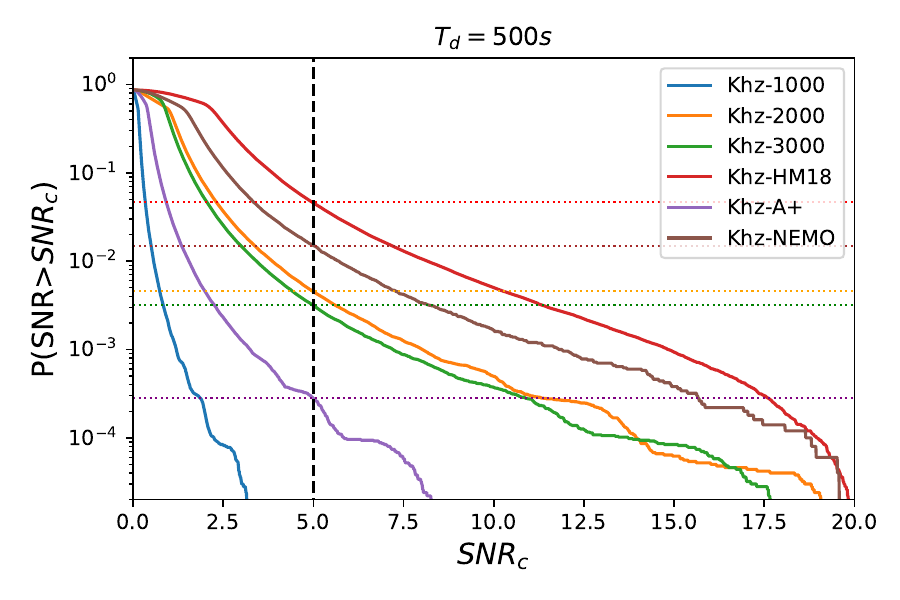}} &
    \resizebox{80mm}{!}{\includegraphics{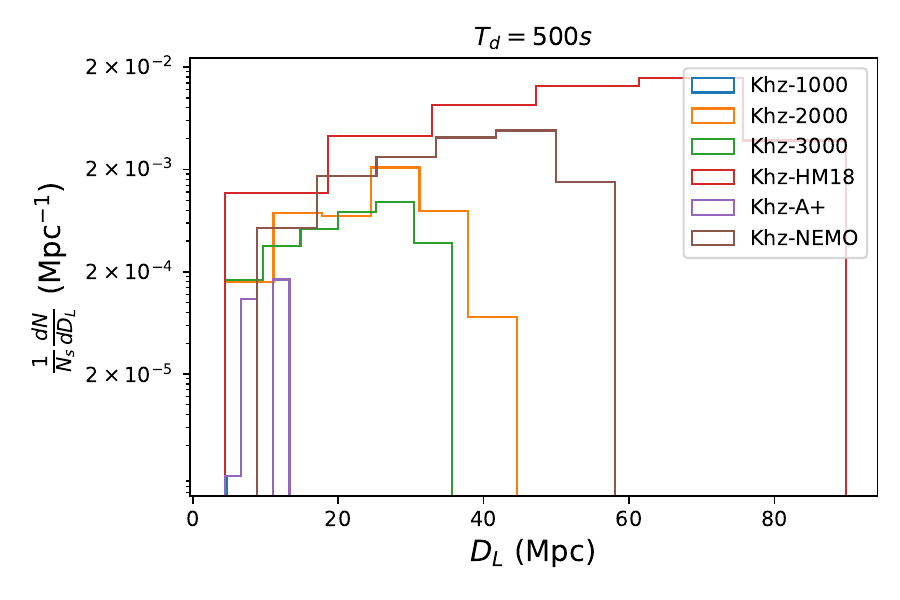}}
    \\
    \resizebox{80mm}{!}{\includegraphics{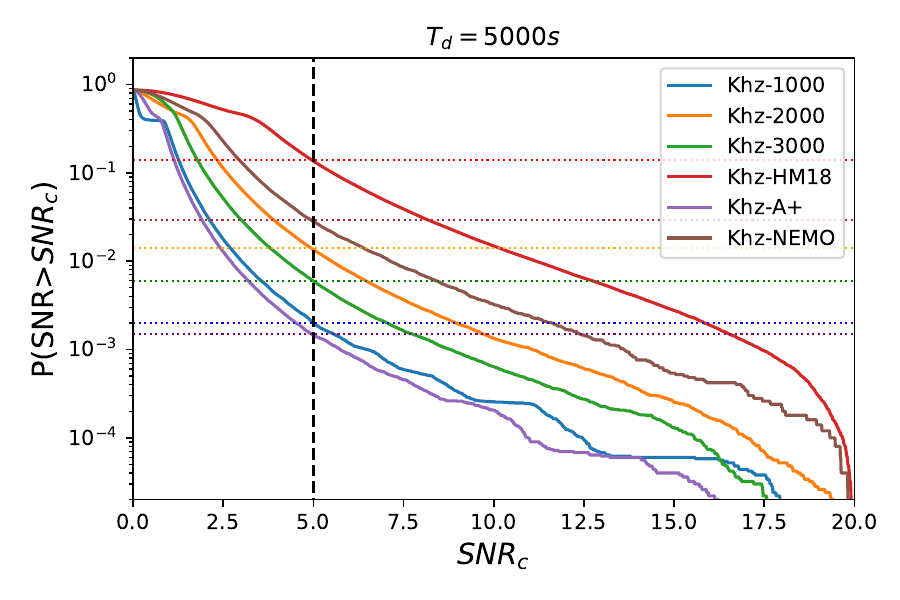}} &
    \resizebox{80mm}{!}{\includegraphics{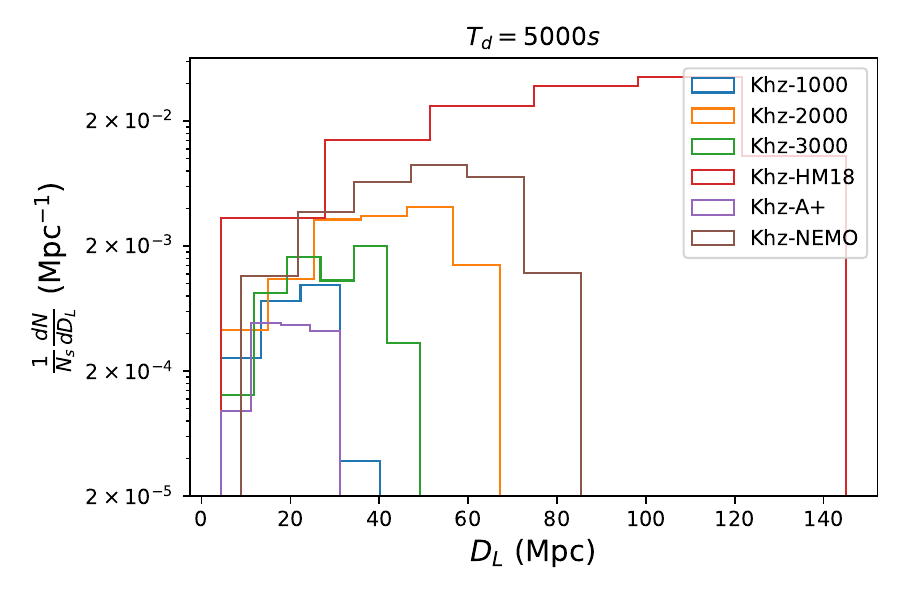}}
    \\
    \resizebox{80mm}{!}{\includegraphics{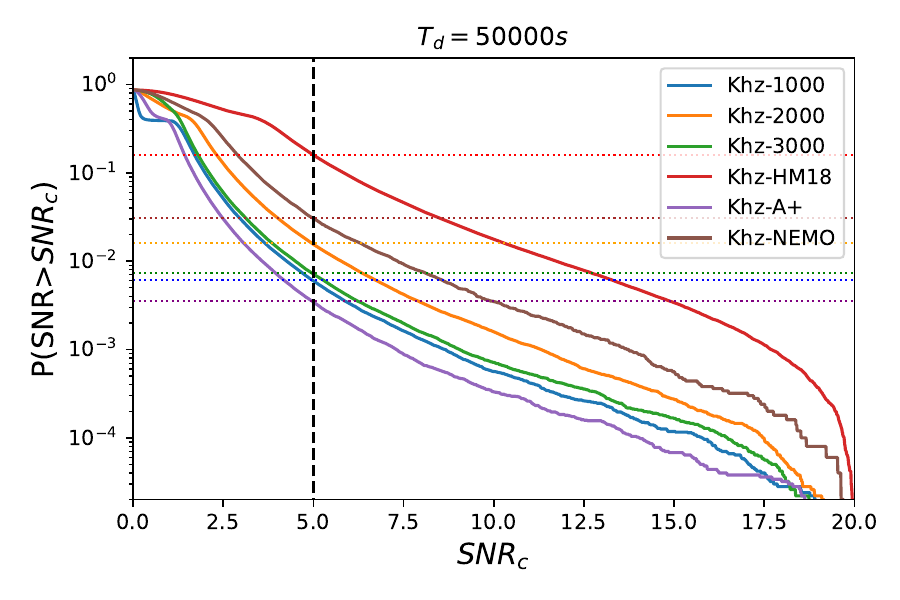}} &
    \resizebox{80mm}{!}{\includegraphics{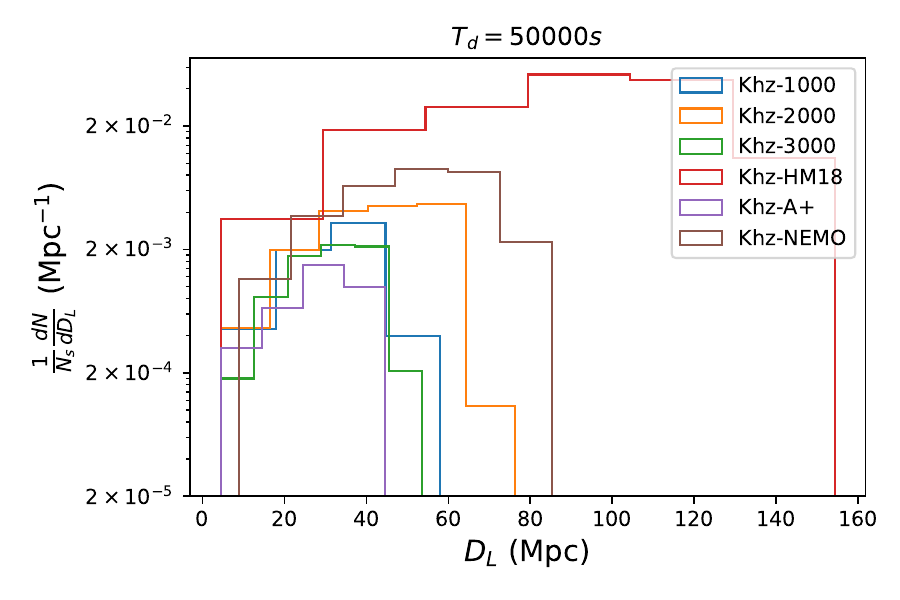}}
    \end{tabular}
    \caption{The left panels show the cumulative SNR distribution for the simulated post-merger GW signals, while the right panels show the distribution of the distances from the GW sources to the observer for those with ${\rm SNR} > 5$. The upper, middle and lower panels show the cases that $T_d = 5 \times 10^2s$, $5 \times 10^3s$, and $5 \times 10^4s$, respectively. Different colors stand for different GW detectors applied. All of these panels is in the case that both GW radiation and EM radiation are considered in the spin-down process.}
    \label{fig:SNR and D_GW and EM}
\end{figure*}

In the case that both the contributions of GW and EM radiation to the spin-down process are not negligible, the fractions of the simulated sources that have a detectable post-merger GW signal are list in Table 3 , while the cumulative distribution of SNRs and the distribution of distances for the detectable post-merger sources are shown in Figure \ref{fig:SNR and D_GW and EM}.

The general trend in this case remains the same as that in the case of GW-dominated spindown, but the fraction with the kilohertz detector designed by \cite{miao2018} have not reached a saturated value. In this case, it is even more unlikely for the secular post-merger GW signals to be detected. For the kilohertz detectors designed with the same techniques as LIGO A+, the fraction can only reach $1.6\%$ even with $T_d = 5 \times 10^4{\rm s}$ while be as low as $0.5\%$ when $T_d = 5\times 10^2{\rm s}$. 

Again, the performance of NEMO is better\citep{nemo} than our designs, which increases fractions to $1.5\%$, $2.9\%$ and $3.1\%$ when $T_d = 5\times 10^2{\rm s}$, $5\times 10^3{\rm s}$ and $5\times 10^4{\rm s}$, respectively. For the kilohertz GW detector designed in \cite{miao2018}, the expected fractions are $4.7\%$, $14\%$ and $16\%$ when $T_d = 5\times 10^2{\rm s}$, $5\times 10^3{\rm s}$ and $5\times 10^4{\rm s}$, respectively.

\begin{table}                                          
\begin{center}{\scriptsize                                          \caption{The fraction of BNS mergers, inside the detecting limit of LIGO-Virgo-KAGRA O4 ($170~{\rm Mpc}$), with a detectable post-merger GW signal. The spin-down process is considered both in GW radiation and EM radiation.}

\begin{tabular}{llll}   
\hline
\hline
&$T_d = 5 \times 10^2~{\rm s}$ &$T_d = 5 \times 10^3~{\rm s}$ & $T_d = 5 \times 10^4~{\rm s}$  \\
\hline
LIGO A+ &0.028\%& 0.15\%&0.35\%  \\
kHz-1000 & 0.00020\%&0.20\% &0.61\%  \\
kHz-2000 &0.46\%&1.4\%&1.6\%    \\
kHz-3000 &0.32\%&0.60\%&0.73\%   \\
NEMO &1.5\%&2.9\%&3.1\%    \\
kHz-HM18 &4.7\%&14\%&16\%    \\
\hline
\hline
\end{tabular} 
\\ \hspace*{\fill} \\

}
\end{center}                                         \label{table:GW_EM}    
\end{table}

\section{Conclusion and discussion}
In this paper, we studied the secular post-merger GW emissions from the newly-formed rigidly-rotating NSs after the BNS mergers. We discussed several kilohertz GW detectors targeting on this type of signals. Then, we simulated the BNS mergers within the detecting limit of LIGO-Virgo-KAGRA O4 ($170{\rm Mpc}$), and infer the fraction of the sources that have a detectable secular post-merger GW signal for each designed kilohertz GW detector, to test their performance.

For the kilohertz GW detector proposed in this paper with only the length of SRC adjusted while other structures remain the same as LIGO A+, the fraction of the sources that have a detectable secular post-merger GW signal is at most $\sim 11\%$ when the length of the effective GW data is $T_d = 5\times 10^4{\rm s}$, and $\sim 1\%$ when $T_d = 5 \times 10^2{\rm s}$, in the GW-dominated spindown case. If the effect of magnetic dipole radiation on the spin-down process is not negligible, this fraction would decrease to $0.5\% - 1.6\%$. These results are based on the detector with peak frequency at $2000 {\rm Hz}$, which is better than that with peak frequency at $1000 {\rm Hz}$ and $3000 {\rm Hz}$, and much better than the case when just searching for secular post-merger GW signals in the data of LIGO A+ itself. According to this fraction, the detecting rate should not be satisfactory. Therefore, new technique needs to be applied to upgrade the detector.

%If the designed $2.5$-generation detector NEMO is used, the sensitivity can be significantly enhanced in a wide frequency range from $1000 {\rm Hz}$ to $3000 {\rm Hz}$, which covers almost all the possible frequencies of the GW emission from a post-merger rigidly rotating NS. 

The proposed $2.5$-generation detector NEMO has much wider sensitive frequency range in kilohertz band, from $1000 {\rm Hz}$ to $3000 {\rm Hz}$, which covers almost all the possible frequencies of the GW emission from a post-merger rigidly rotating NS. It it was used, the fraction of the sources that followed by a detectable secular post-merger GW signal would reach $\sim 42\%$ if the spin-down processes for the post-merger rigidly rotating NSs are all dominated by GW radiation and $T_d > 5\times 10^4{\rm s}$, while the fraction decreases to $\sim 5\%$ when $T_d = 5 \times 10^2$. If magnetic dipole radiation also contributes to the spindown of the post-merger rigidly rotating NSs, this fraction decrease to $\sim 3\%$ when $T_d = 5\times 10^3 - 5\times 10^4 {\rm s}$ and to $\sim 1.5\%$ when $T_d = 5\times 10^2 {\rm s}$. Furthermore, if more quantum technologies were applied to GW detectors as proposed by \cite{miao2018}, the sensitivity of the GW detector and the results obtained will be even better. The fraction of the sources that followed by a detectable secular post-merger GW signal would reach $\sim 45\%$ if the spin-down processes for the post-merger rigidly rotating NSs are all dominated by GW radiation and $T_d > 5\times 10^3{\rm s}$, while the fraction decreases to $\sim 11\%$ when $T_d = 5 \times 10^2$. If magnetic dipole radiation also contributes to the spindown of the post-merger rigidly rotating NSs, this fraction decrease to $\sim 15\%$ when $T_d = 5\times 10^3 - 5\times 10^4 {\rm s}$ and to $\sim 5\%$ when $T_d = 5\times 10^2 {\rm s}$. 

There are also other design schemes to obtain high-frequency gravitational wave detectors at the cost of decreasing the sensitivity. For example, detuning the phase of SRC directly without changing the length of SRC can easily change the peak frequency of detector by adjusting the phase of signal recycling mirror\citep{detune}. However, the power spectrum density of this detector in peak frequency will be twice than the detector proposed in this paper in theoretical because of losing half of the sideband. For another example, we can using an auxiliary resonant cavity with maintaining the same length of SRC. This auxiliary resonant cavity is used to compensate for phase to get the peak frequency we want. Of course, using an auxiliary resonant cavity will definitely bring new losses and noises and lead to the decreasing of sensitivity. The advantage of this design is that we can mainly adjust the transmission of the resonant cavity, thereby controlling its scale within a certain space. Besides, another way to improve the sensitivity of the GW detector is to use longer arm length. longer length of the arm cavity means the higher power in the main interferometer. It can improve the sensitivity of the detector in all frequency bands, not just in the kilohertz band. In this case, the antenna response is needed to be considered. It will decrease the SNR of the GW detector.  Of course, it is also a very interesting subject to know the ultimate light intensity that the interferometer can withstand.

In addtion, we compared the kilohertz detectors discussed in this paper with future ground-base GW detectors. We find that, targeting on this specific kind of sources, the performance of the kilohertz detectors designed with the same technique used for LIGO A+ could be comparable to LIGO Voyager, while the kilohertz detector with more quantum technologies (i.e. \cite{miao2018}) could be comparable to the third-generation GW detectors (i.e. Einstein Telescope and Cosmic Explorer). With Einstein Telescope and Cosmic Explorer, we will definitely detect more of such signals, but the fraction of the sources that have a detectable secular post-merger GW signal may not necessarily increase due to the highly enhanced detecting limit for the inspiral signal of BNS mergers. 

In this work, we did not consider the inclination angle of the post-merger NSs, which might make the characteristic strains of the GW radiation slightly overestimated. Therefore, the estimated fraction of the sources that have a detectable secular post-merger GW signal could be even lower. We treat the $\epsilon$ value as a constant during the whole spin-down process because its evolution over time is highly model-dependent thus unclear. Therefore, the $\epsilon$ value in this work should be an effective value. Our results could be refined if the evolution of $\epsilon$ is well determined in the future.

\section*{Acknowledgements}
We are grateful to Xilong Fan for constructive discussion. This work was supported by the National Natural Science Foundation of China under Grants Nos. 12021003, 11920101003 and 11633001, the Strategic Priority Research Program of the Chinese Academy of Sciences under Grant No. XDB23000000, the Interdiscipline Research Funds of Beijing Normal University, the Shanghai Post-doctoral Excellence Program under Grant No. 2022671, and the China Postdoctoral Science Foundation under Grant No. 2023M732713.

%%%%%%%%%%%%%%%%%%%%%%%%%%%%%%%%%%%%%%%%%%%%%%%%%%
\section*{Data Availability}
The BAT and XRT light curve data for SGRBs are obtained by using the public data from the Swift archive: \url{https://www.swift.ac.uk/burst\_analyser/}, which is provided by the Swift Data Center.



%\bibliographystyle{mnras}
%\bibliography{mybib} % if your bibtex file is called example.bib

%\begin{thebibliography}{99}
%\end{thebibliography}

%%%%%%%%%%%%%%%%%%%%%%%%%%%%%%%%%%%%%%%%%%%%%%%%%%

%%%%%%%%%%%%%%%%% APPENDICES %%%%%%%%%%%%%%%%%%%%%

\appendix
\section{X-ray internal plateau}
We conducted an analysis of 144 short-duration gamma-ray bursts (SGRBs) that were observed with the Swift satellite between January 2005 and January 2023. Our aim was to investigate the presence of plateau features in the light curves of their X-ray afterglows, which could indicate the collapse of a supra-massive neutron star (NS) into a black hole (BH). We extrapolated the available BAT-band data (15-150 keV) to XRT-band (0.3--10 keV) using a single power-law spectrum. We then applied a temporal fit to the combined light curve, utilizing a smooth broken power-law function, in order to identify potential plateau features. These features are characterized by shallow decay slopes smaller than 0.5, followed by a steep decay index greater than 3. Our analysis led to the identification of 35 SGRBs displaying internal plateau characteristics, constituting approximately 24\% of the total sGRB sample. A detailed list of the 35 SGRBs with internal plateau characteristics, and their corresponding breaking times, can be found in Table A1.

\begin{table}
\caption{breaking times of X-ray internal plateaus}
\centering
\begin{tabular}{|c|c|c|c|}
  \hline
SGRB & breaking time (${\rm s}$) & SGRB & breaking time (${\rm s}$) \\\hline
GRB 050724 & 139 & GRB 210726A & 136\\ \hline
GRB 051210 & 67 & GRB 200907B & 426\\ \hline
GRB 051227 & 89 & GRB 200411A & 144\\ \hline
GRB 060801 & 212 & GRB 200219A & 202\\ \hline
GRB 061006 & 99  &   GRB 191031D & 11\\ \hline
GRB 061201 & 2223  & GRB 181123B & 471\\ \hline
GRB 070714B & 82  &  GRB 180805B & 166\\ \hline
GRB 070724A & 77 &  GRB 180727A & 29 \\ \hline
GRB 071227 & 69  &  GRB 180402A & 136\\ \hline
GRB 080702A & 586 & GRB 160821B & 180\\ \hline
GRB 080905A & 13 & GRB 160624A & 158\\ \hline
GRB 080919 & 340 & GRB 150423A & 3710\\ \hline
GRB 081024A & 102 & GRB 150301A & 114\\ \hline
GRB 090510 & 1494 & GRB 120521A & 270\\ \hline
GRB 090515 & 178 & GRB 120305A & 188\\ \hline
GRB 100117A & 252 & GRB 111121A & 56\\ \hline
GRB 100625A & 200 & GRB 101219A & 197\\ \hline
GRB 100702A & 201 \\ \hline

\end{tabular}
\label{sec:breaking times}
\end{table}
%%%%%%%%%%%%%%%%%%%%%%%%%%%%%%%%%%%%%%%%%%%%%%%%%%

% Don't change these lines
\bsp	% typesetting comment
\label{lastpage}
\end{document}